\documentclass{article}
\usepackage{epsfig}
\usepackage{amsmath}
\usepackage{amssymb}


\begin{document}

\author{Sankhasubhra Nag\thanks{Saha Institute of Nuclear Physics, 1-AF 
Bidhannagar, Kolkata 700 064}, Avijit Lahiri\thanks{Vidyasagar Evening 
College, Kolkata 700 006} and Gautam Ghosh\thanks{Saha Institute of Nuclear 
Physics, 1-AF Bidhannagar, Kolkata 700 064, Email:gghosh@tnp.saha.ernet.in}}

\title{Entropy production due to coupling to a heat bath in the kicked rotor 
problem}

\date{\today} 

\maketitle

\begin{abstract}Considering a kicked rotor coupled to a model heat bath both 
the classical and quantum entropy productions are calculated exactly. Starting
with an initial wave packet, the von Neuman entropy as a function of time 
is determined from the reduced density matrix while the Liouville evolution
of the corresponding Husimi distribution provides us with the classical 
entropy. It is found that both these entropies agree reasonably satisfying
the same asymptotic growth law and more importantly both are proportional 
to the classical Liapounov exponent.  \end{abstract}

keywords: quantum open system, entropy


The question of quantum classical correspondence in the context of open
systems is an interesting one. It has been argued, for instance by
Zurek and Paz\cite{1z}, that  even a weak interaction with a random environment
leads to the loss of quantum coherence and that classical behaviour may then
be an emergent  property of such systems. In other words, environmental
effects may have a significant role to play in  quantum classical
correspondence in general and in the quantum dynamics of classically chaotic
systems in particular. The von Neumann entropy production of an unstable
oscillator interacting with a heat bath has been shown \cite{1z} to depend
monotonically on the classical  local Lyapunov exponent. A more extensive
study of the kicked rotor in a bath by Miller and Sarkar \cite{2m} brings out
in greater detail this correspondence between quantum entropy production and
classical Lyapunov exponent. The quantum evolution of classically chaotic open
systems has been shown to possess a sensitive dependence on initial conditions
\cite{3t} and provides another  interesting indication of the role
of the environment. As a further instance     one may mention the high degree
of sensitivity of quantum dynamic localization \cite{4c} to the presence of
noise found in a study on the kicked rotor \cite{5o}.Extensive studies of the 
kicked rotor coupled to a bath were carried out
 by Dittrich and Graham\cite{6d} and by Cohen\cite{7c}. These authors were interested in 
reducing the problem and showing its equivalence to a classical stochastic
map in the semiclassical limit with the aid of the Wigner function 
representation of the density matrix.

 In the present paper we look into a few aspects of quantum classical 
correspondence in open systems by focussing on the kicked rotor interacting
with a bath of harmonic oscillators through what is termed a nondemolition 
coupling. The latter involves an interaction Hamiltonian commuting with the
system Hamiltonian as a result of which the reduced density matrix can be
obtained exactly \cite{6s}. This feature has been used in \cite{2m} in
computing the von Neumann entropy production rate of the kicked rotor. We show
that the special     nature of the coupling allows one to obtain the classical
reduced             distribution function \cite{7w} as well and then to
compare the diffusion rate and the entropy production at the classical and
quantum levels under similar       initial conditions. Numerical computations
show that in the semiclassical limit the asymptotic
von Neuman entropy follows the same classical growth law as obtains for a 
diffusion process viz. $s(t)=A+B\ln t$ and that this regime is established
smoothly after a very few kicks even for relatively weak bath couplings.
Of more interest is the fact that in the asymptotic region the quantum
 entropy is proportional to the classical Liapounov exponent which may
 perhaps be looked upon as a signature of quantum chaos in semiclassical
 quantum mechanics.

      We take a cylindrical phase space ($0 < q \leq 2\pi; -\infty <  p
< \infty$) and define the kicked rotor Hamiltonian to be,
\begin{equation} H_{rotor}=p^2/2 + K \cos
q\sum_{n=-\infty}^{+\infty}\delta(t-n).  \end{equation}The classical dynamics
is defined by the standard map,  
\begin{eqnarray}  p_{n+1}&=&p_n+K\sin q_{n+1}, \label{pmap}\\  
q_{n+1}&=&q_n+p_n\;\;\;\;\;\;\; \;\;mod\: 2\pi,  \label{qmap} 
\end{eqnarray} where $q_n$ and $p_n$ are the
values just after the $n$th kick. Below a critical kick strength $K_c (\sim
.97)$ KAM tori prevent the indefinite growth in energy while above $K_c$, one
has a mixed phase space of islands  surrounded by a
chaotic sea and the diffusive growth law holds viz. \(\langle
(p_n-p_0)^2\rangle=D(K)n,  \) where the average is over a
distribution of initial points and the diffusion coefficient tends to $K^2/2$
for large $K$.  In quantum mechanics
the evolution is governed by the single step unitary evolution operator which
takes the state from just after the $n$th kick to just after the $(n+1)$th kick
and is given by, \begin{eqnarray}
U(n+1,n)\equiv U_k U_f=\exp[{iK \over \hbar}\cos
q]\exp[-ip^2/(2\hbar)]. \label{qevol} \end{eqnarray}

  In the momentum representation $U_f$ is just a phase factor and the matrix 
element for $U_k$ is given by,  \begin{equation} \langle l\mid U_k\mid
m\rangle =(i)^{l-m}J_{l-m}(K/\hbar). \end{equation}   Numerical computation
with an initial wave packet shows a diffusive growth in energy for a finite
time determined by the values of $K$ and $\hbar$ after which the energy
oscillates quasiperiodically about an average value \cite{4c,8i}. 

         The total Hamiltonian for a system with a nondemolition coupling to
an  oscillator bath is chosen to be, \begin{eqnarray}H &=& H_0(p)+{1\over
2}\sum_k(p_k^2+\omega_k^2q_k^2)\nonumber\\& &+\phi(H_0)\sum_kc_k\sqrt{
\omega_k}q_k+{1\over 2}\sum_k{c_k^2\over\omega_k}\phi^2(H_0),\label{coupH}
\end{eqnarray} where the first two terms are the system and bath Hamiltonians
respectively. The third term is the interaction with $c_k$ being the coupling
constant to the $k$th mode, $\phi$ being an arbitrarily chosen function, while
the fourth one is a renormalization term.  We will work out the time
evolution for a general Hamiltonian $H_0(p)$ and then specialize to the case
of the rotor. Notice that the interaction is a function of the system
Hamiltonian and therefore momentum will be conserved. Making a canonical
transformation with the generating function,  
\begin{equation}F(q,q_k;P,P_k)=qP+\sum_kq_kP_k+\phi(P)\sum_k{c_k\over
\omega_k^{3/2}}P_k,\end{equation}   the Hamiltonian is reduced to the
uncoupled form, \begin{equation}H=H_0(P)+{1\over
2}\sum_k(P_k^2+\omega_k^2Q_k^2). \label{uncoupH} \end{equation}  Solving the
equations of motion obtained from \ (\ref{uncoupH}) and reverting back to
original variables we get our required  solutions viz. $ q(0) \text{ and }
p(0)$ in terms of $q, p,q_k(0) \text{ and } p_k(0)$.    As an initial
point $\{q(0),p(0)\}$ evolves in time $t$ to $\{q,p\}$ the  distribution
function evolves from $f^0[q(0),p(0)]$ to $f[q,p]$ maintaining its value at
the corresponding points. However, since the evolution depends on the bath
variables, the reduced distribution function is obtained by integrating  over
the bath quantities which are assumed initially to have values pertaining to
thermal equilibrium. Thus, \begin{equation}f_{red.}(q,p,t)=\int
f^0[q(0),p(0)]\prod_k{g_k\over z_k}dp_k(0)dq_k(0),
\label{crfred}\end{equation} where in the r.h.s. $[q(0),p(0)]$ are understood
to be written in terms of $[q,p]$. The factor $g_k$ is the thermal probability
for the $k$th bath mode at temperature $T(=1/\beta)$ and it is normalized by
the partition function $z_k$.   We carry out the integration in equation\
(\ref{crfred}) and go to the continuum limit  whereby the sum over modes is
replaced by an integration over frequency with an ohmic spectral density
function given by, \begin{equation}\sum_k\rightarrow\int f(\omega)d\omega,    
      ~~~~f(\omega)=\eta {\omega\over c^2(\omega)} e^{-{\omega / \omega_c}}. 
\end{equation} The reduced distribution function is given by, \begin{eqnarray}
f_{red.}(q,p,t)&=&\sqrt{\frac{\beta }{2\pi s(t)}}\int drf^{0}[q-({\partial H_0
\over \partial p})t+r\nonumber\\&~&-\phi\phi^{\prime}\eta \arctan \omega
_{c}t,p]\exp (-\frac{\beta r^{2}}{2s(t)}),\label{fredf}\\ s(t)&=&2\eta
{\phi^{\prime}}^{2}[t\arctan \omega _{c}t-\frac{1}{2\omega _{c}}\ln (1+\omega
_{c}^{2}t^{2})].\end{eqnarray}If we look at the fundamental  solution of \
(\ref{fredf}) when \(f^0\) is a \(\delta\)-function we find \(f_{red.}\) to
be a gaussian . For \(\omega_ct\ll1,\) \(s(t)\propto t^2\) and the gaussian
spreads fast but after some time depending on \(\omega_c \),
\(\arctan\omega_ct\) saturates  at \( \pi/2\) and \(s(t)\propto t\), so that
equation\ (\ref{fredf}) describes a normal diffusion process. In the case of
the rotor \( (\partial H_0(p)/\partial p)=p \), and for the  coupling we take
\( \phi(p)=p \), so that \( \phi^{\prime}=1 \). The fourth term in\
(\ref{coupH}) therefore renormalizes the rotor mass. Because of the
nondemolition nature of the coupling the $p$-distribution remains the same  in
the absence of the kicking term whereas the $q$-distribution suffers a shift
and a gaussian local averaging. However, as it turns out, the bath brings
about a mixing in the $p$-distribution on a fine scale when the kicks couple
$q$ and $p$. . 

 For the quantized form of the Hamiltonian\ (\ref{coupH})  the matrix elements
of the reduced density operator for the rotor at time $t$  are given by
\cite{2m,6s}, \begin{eqnarray}\langle
m\mid\rho_{red.}(t)\mid n \rangle&=&\exp[-i\hbar(m^2-n^2)t/2] \nonumber\\
&&\langle m\mid\rho(0)\mid n \rangle
\:\exp[-iA^1_{mn}-A^2_{mn}],\label{rhored}\end{eqnarray}   where, 
\begin{eqnarray}A^1_{mn}&=&\eta \hbar(m^2-n^2)\tan^{-1}\omega_ct,
\\A^2_{mn}&=&{\eta\hbar\over 2}(m-n)^2
\ln(1+\omega_c^2t^2)\nonumber\\&&+\eta\hbar(m-n)^2\ln\prod_{k=1}^{\infty}
\{1+({\omega_ct\over 1+k\beta\omega_c})^2\}.\end{eqnarray}   Notice
that the diagonal matrix elements are unaffected by the interaction with the
bath when the kicking term is absent.

In analysing the quantum evolution we take an initial normalized wave packet
of the form, $\mid\psi\rangle =\sum_la_l\mid l\rangle $  which is localized
and peaked around the momentum $p=\hbar l_0$. We achieve similar initial
conditions for the classical evolution by taking the Husimi distribution
\cite{9ta} corresponding to $\mid\psi\rangle$. This is given by,
\begin{eqnarray}f(q,p)\equiv \mid \langle q,p\mid
\psi\rangle\mid^2&=&\sum_{m,n}{a_m^* a_n\over
\sqrt{\pi}}\exp[-p^2/\hbar^2\nonumber\\+(m+n)p/\hbar&-&(m^2+n^2)/2]\cos(m-n)q
, \label{husimi}\end{eqnarray}which is a diagonal approximation of the density
matrix in the coherent state   representation. $f(q,p)$ is periodic in $q$
with period $2\pi$ so that\ (\ref{husimi}) is     really defined on the
cylinder.  The coefficients $a_l$ are chosen such that $\mid\psi\rangle$ 
represents a  wave packet with $\Delta p=\Delta q=\hbar/2$. This is achieved by
parametrising $a_l$ in the form,  $ a_l=N\exp[-a l^2+bl], $ where $a$ is real
and $b$ is complex such that the three parameters $a$ and the real and
imaginary parts of $b$ determine the centre and the width of the wave packet.

For the quantum evolution the computation starts with the density matrix
corresponding to $\mid\psi\rangle$. After  unit time step the reduced density
matrix is evaluated using equation\ (\ref{rhored}). The unitary kick operator
$U_k$ then connects the matrix elements for the  density  operators just
before and after the kick according to (4). The  von
Neumann entropy is calculated by using the definition, \begin{equation}
S\equiv -Tr\rho \ln\rho =-\sum_i\lambda_i\ln\lambda_i,   \end{equation}where
the $\lambda_i$ are the eigenvalues of the density operator. The  computation
is repeated till the entropy reaches its asymptotic growth rate.   To compute
the classical evolution we start with the Husimi distribution  corresponding
to the initial quantum wave packet and use equation\ (\ref{crfred}) to find
the reduced distribution after unit time step. The distributions just before 
and after the kicks are related by a simple shift in argument viz. 
$f_{red.}^+(q,p)=f_{red.}^-(q,p-K\sin q).$  The entropy is calculated by using
the formula,\begin{equation} S_H=-\int f\ln f {dp dq \over 2\pi\hbar},  
~~~~~~~~~~~~\int f {dp dq\over 2\pi
\hbar} = 1.\label{Sh} \end{equation}  It is known that \(S_H \) calculated from
$f(q,p)$ as defined in\ (\ref{husimi}) satisfies the inequality, \(S_H \geq
S.\) It turns out that the classical approximation is good as long as \(f(q,p)
\) is a smooth function spread over an area in phase space that is much
greater than \(2\pi\hbar \). If there are small distance fluctuations or if \(
f(q,p) \) is concentrated on a small region of the phase space, then the
classical approximation can be very bad \cite{7w}. This fact will be borne out
in our numerical computations.  Notice that the von Neumann entropy for the
initial pure state is zero. However, this is not true of\ (\ref{Sh}). The
Husimi distribution corresponding to the initial wave packet is sufficiently
well localised in $q$ to be considered a gaussian with \(\Delta p \Delta q =
\hbar\) and its entropy is just unity.The distribution is chosen to be centred
at \(q=0, p =\pi\hbar \) which lies in a chaotic region for all the
\(K\)-values cosidered by us. 

\begin{figure} 
\hskip 0.5cm
\epsfig{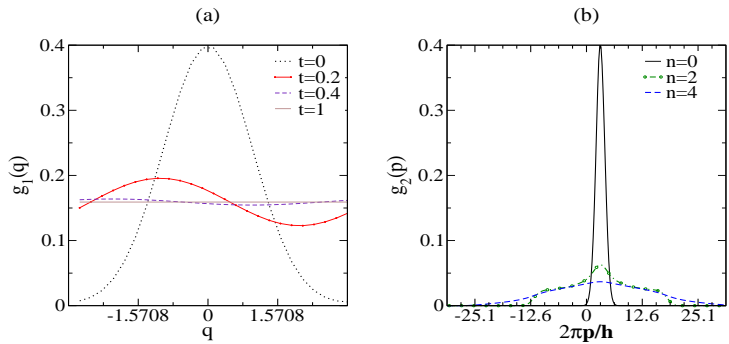}
\hskip 0.3cm
\caption{Evolution of (a) $g_1(q)$ and (b) $g_2(p)$ for $\eta=1.0,
K=3.5$.}  
\label{fig1} 
\end{figure}

To understand the diffusion process in detail we compute the $q$-distribution,
$g_1 (q)$ and the $p$-distribution, $g_2(p)$ as the integrals of $f(q,p)$ over
$p$ and $q$ respectively. The diffusion rate in \(q\) is controlled by the
parameter \( \omega_c \) and \( \beta \). For our choice, \( \omega_c
=5/\hbar\approx 10\text{ and }\beta=0.1\) satisfying $\beta\hbar\omega_c<1$, we
find that \( g_1(q) \) almost reaches the value \(({1 \over 2\pi})\) before the
first kick [fig.\ (\ref{fig1}a)]. Decreasing \( \omega_c \), will increase the
equilibriation time for the \(q\)-distribution. The large scale structure of
the \(p\)-distribution on the other hand is determined by the kicks, the bath
providing a mixing on a finer scale through the coupling with the kicks.
Figure\ (\ref{fig1}b) shows the initial evolution of \(g_2(p)\).

\begin{figure}
\hskip 2.0cm
\epsfig {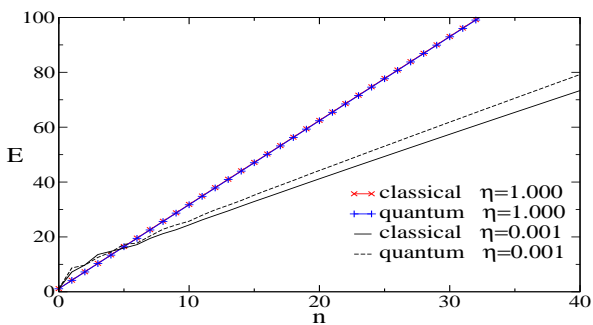}
\caption{Comparison of  classical and quantum energy growth for
$K=3.5,\hbar=0.46$.} 
\label{fig2}
\end{figure}

Next, we compute the classical and quantum energies. The effect of $\hbar$
for fixed $K$ (the relevant parameter actually is  $K/\hbar$) on the variation
of $\langle E\rangle$ with $t$ as also on the entropy  production (see below;
neither of these two is shown in our figures) indicates  that one can
distinguish between a semiclassical and strongly quantum regime: while
oscillations due to quantum correlations are pronounced in the latter, they
are almost absent in the former. All our numerical work relates mainly  to the
semiclassical regime with $\hbar$ taken as a rational multiple of the golden
mean $(\sqrt 5-1)/2$ to avoid resonances. The energy growth is typically
diffusive for both  classical and quantum dynamics and the diffusion rate
increases with $K$. What is interesting is the effect of the bath coupling
strength $\eta$ on the energy growth\ (fig.\ref{fig2}): on increasing $\eta$
from a low value one finds that the classical and quantum energy growth curves
quickly approach each other (cf. Ott, Antonsen and Hanson \cite{5o}) and their growth
rates saturate at $K^2/4$ even for comparatively low values of $K$. This is
consistent with the role  of the bath outlined above.

\begin{figure}
\hskip 2cm
\epsfig {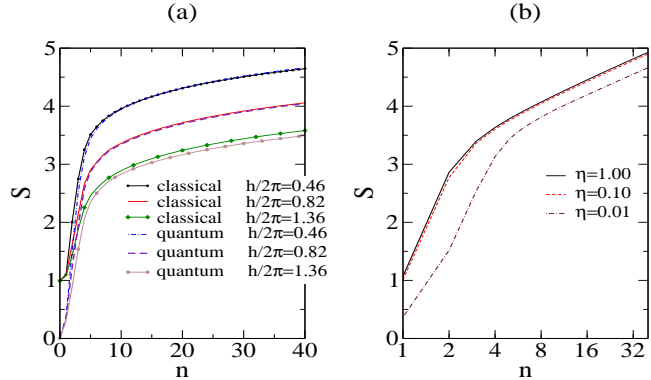}
\caption{Entropy growth in the quantum case for (a) different $\hbar$ with
$K=3.5, \eta=10^{-2}$ and (b) different $\eta$ with $K=3.5,\hbar=0.46$.}
\label{fig3} \end{figure}

Figures (\ref{fig3}(a,b)) present results on entropy production. Note that
there is an $\hbar$ dependence in the classical entropy $S_H$ arising from the
Husimi density corresponding to an initial wave packet. The evolution of the
Husimi density satisfies the Liuville equation in the lowest order in $\hbar$
and that is what we have considered here. The classical and quantum
entropy  productions are compared in fig. \ref{fig3}(a) which shows that the
two converges quickly excepting for comparatively large values of $\hbar$ and
even when they differ their asymptotic growth rates agree. The time dependence
of entropy production is elucidated in fig. \ref{fig3}(b) which uses a
logarithmic time scale: asymptotically the entropy settles down to a
logarithmic growth. Curves for different  values of $\eta$ once again show a
quick saturation with increasing $\eta$.     
 
The asymptotic entropy production fits nicely with the formula $S=A+B\ln n$
which we now explain by referring to the classical entropy $S_H$. Entropy
production occurs due to the coarse graining provided by the bath along with
the diffusion provided by the kicks. The bath quickly uniformizes the
q-distribution  and converts the p-distribution to a smoothed gaussian so that
in the asymptotic regime  \[f(q,p)={\hbar\over {\sqrt {2\pi}\Delta
p}}exp{(-{p^2 \over {2\Delta p^2}})}.\]Together with the diffusive growth law,
 $\Delta p=(K/\surd 2)n^{1\over 2}$ this yields for the entropy,
\begin{equation}S_H={1\over 2}+\ln({\sqrt{\pi}\over\hbar}) +\ln K +{1\over
2}\ln n.\label{Shest}\end{equation} Since for large $K$, the Lyapunov exponent
is, $\lambda=\ln{K\over 2}$, the coefficient $A$ is linear in the Lyapunov
exponent for large $K$.

\begin{figure}
\epsfig {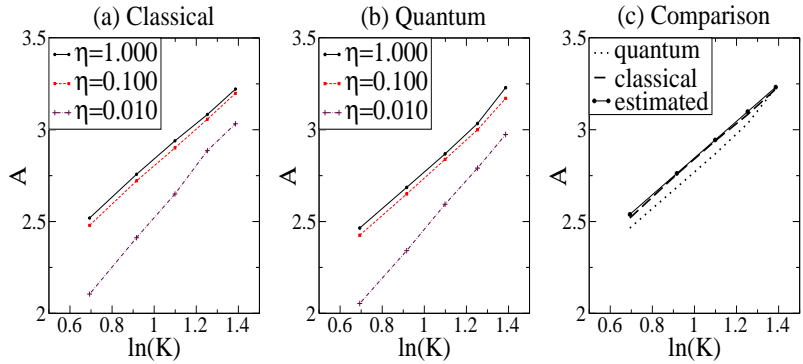}
\vskip 0.3cm
\caption{Variation of $A$ with $\ln K$ for different $\eta$ in  the
(a) classical and (b) quantum case. (c) shows a comparison with the values
estimated from\ (\ref{Shest}) for $\eta=1.0$.}  \label{fig4} \end{figure}

The growth law $S=A+B\ln n$ with $A$ depending linearly on $\ln K$ and
$B=1/2 $  is numerically found to apply to the quantum entropy as well in the
 semiclassical regime. Figures \ref{fig4}(a,b) depicts the variation of $A$
with $\ln K$ as computed numerically from the classical and quantum entropy
production data where the linear dependence is evident and where the role of
$\eta$ in quantum classical correspondence is apparent once again. Figure
\ref{fig4}(c) compares the numerically computed values of $A$ with the
estimated values from equation\ (\ref{Shest}).
 
\begin{figure}
\hskip 2.0cm
\epsfig{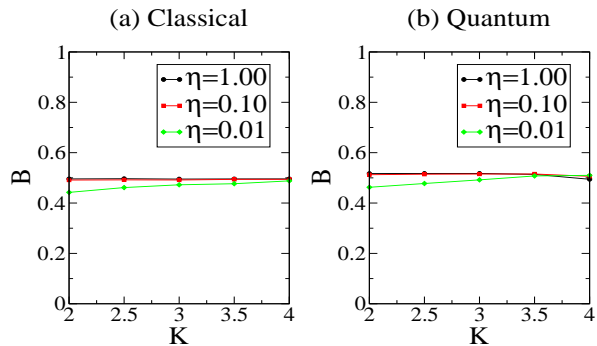}
\caption{Plot of $B$ with $K$ for different $\eta$ in (a)classical case and
(b)quantum case.} \label{fig5}
\end{figure}

Figures \ref{fig5}(a,b) present corresponding data for the coefficient $B$
which is indeed found to be close to $1/2$. The discrepancy is explained by the
presence of residual phase space barriers to diffusion as also incomplete
mixing which disappears more and more with increasing $K$ and $\eta$. In
figures \ref{fig4} and \ref{fig5}, $\hbar$ is held at $0.46$.

In conclusion, the findings presented above underline the   role of the bath
in establishing a  quantum-classical correspondence wherein the quantum entropy
production carries with it the characteristics of the  classical phase space
structure: for instance, the quantum entropy is tied to the classical Lyapunov
exponent  \cite{1z,2m}. It would be interesting to look into the entropy
production in the kicked Harper model where the phase space is compact and the
diffusive growth law $(\Delta p\propto\sqrt n)$ does not hold.


\begin{thebibliography}{0} 
\bibitem[1]{1z} W. H. Zurek and J. P. Paz, Phys. Rev. 
Lett.72, 2508 (1994); see however, G. Casati and B.V. Chirikov, Phys. Rev. 
Lett.75, 350 (1995) for an alternate view on the role of external noise.
\bibitem[2]{2m}P. A. Miller and S. Sarkar, Nonlinearity 12, 419 (1999). 
\bibitem[3]{3t} M. Toda,S. Adachi and K. Ikeda, Prog. Theo. Phys. Suppl. 98, 
 323 (1989).  
\bibitem[4]{4c}G. Casati, B. V. Chirikov, J. Ford and F. M. Izrailev, Lecture 
Notes in Physics
93, 334 (1979). 
\bibitem[5]{5o} E. Ott, T. M. Antonsen, Jr. and J. D. Hanson,
Phys. Rev. Lett. 53, 2187  (1984).
\bibitem[6]{6d}T. Dittrich and R. Graham, Ann. Phys. 200 (1990) 363.
\bibitem[7]{7c}D. Cohen, J. Phys. A 27 (1994) 4805 and references contained 
therein. 
\bibitem[8]{6s} J. Shao, M. Ge and H. Cheng,
Phys. Rev. E 53, 1243 (1996). 
\bibitem[9]{7w} A. Wehrl, Rev. Mod. Phys. 50, 221
(1978). 
\bibitem[10]{8i} F. M. Izrailev, Phys. Rep. 196, 300 (1990).
\bibitem[11]{9ta} K. Takahashi, Prog. Theo. Phys. Suppl. 98, 109
(1999).  
\end{thebibliography}
\end{document}